\documentclass[journal]{IEEEtran}

\ifCLASSINFOpdf
\else
\fi

\usepackage{array}
\usepackage{amsmath}
\usepackage{graphicx}
\usepackage{subcaption}
\usepackage{amssymb} 
\usepackage{booktabs}
\usepackage{dblfloatfix}
\usepackage{tabularx}

\usepackage{makecell}
\usepackage{wasysym}
\usepackage{threeparttable}
\usepackage{hyperref} 
\newcommand{\yes}{\ensuremath{\checkmark}}
\newcommand{\partialyes}{\LEFTcircle}
\newcommand{\no}{--}
\usepackage{cite}
\newcolumntype{Y}{>{\centering\arraybackslash}X}
\begin{document}

\title{Scalable Reservoir Computing for Disturbance Localization and Identification in Power Grids}

\author{Erin~Obermayer,
    Amir~Sajadi,~\IEEEmembership{Senior~Member,~IEEE,} \\
    Bri-Mathias~Hodge,~\IEEEmembership{Senior Member,~IEEE,} and
    Juan~G.~Restrepo  

\thanks{This work was supported in part by the NSF under Grant DMS-2205967.}
\thanks{The authors are with the University of Colorado Boulder, Boulder, CO 80309 USA (corresponding author: Erin.Obermayer@colorado.edu).}

}


\maketitle
\begin{abstract}
In this paper, we present a novel method for the localization and temporal waveform reconstruction of disturbances in complex power grids using observations of system state variables. Although various methods exist for detecting these events, accurately identifying and responding to them remains a significant challenge due to the grid's size, speed, complexity, and inherently dynamic behavior. We formulate a method for disturbance localization and reconstruction based on reservoir computing, a recurrent neural network architecture. We show that our method is able to recover the time dependence of previously unobserved disturbances, is robust to missing information, and can operate online. The approach requires only observed system data and knowledge of the network topology, without requiring a physical model of the system dynamics, and can recover simultaneous disturbances occurring at multiple locations. We implement a scalable parallelized reservoir architecture and demonstrate our algorithm on test IEEE cases and a reduced WECC system.
\end{abstract}


\IEEEpeerreviewmaketitle

\vspace{-1em}

\section{Introduction}\label{intro}

\IEEEPARstart{P}{ower} grids are the backbone of modern society. Any interruptions in power delivery can disrupt essential services and pose serious risks to economic activity, public safety, and critical infrastructure, underscoring the lessons learned from the recent power outages in Texas \cite{texasFreeze2021}. 
This critical infrastructure remains vulnerable to external disturbances that can result in abnormal conditions, topologies, or operating states. Failure to detect and localize these events quickly bars the operators from taking mitigation actions which could escalate the abnormalities from local events into network-wide challenges, culminating in a widespread blackout, as recently occurred in Spain in 2025 \cite{spainBlackout} or the 3.8 GW load disconnection event in Virginia in 2026 \cite{3p8utilitydive}. 

A review of the literature suggests that a wide range of localization methods have been developed using wide-area measurement systems, synchrophasor data, optimization, and machine learning techniques. Early work exploited the propagation of electromechanical waves 
to estimate disturbance sources \cite{you}, while subsequent studies proposed data-driven optimization methods \cite{lee}, distributed online localization algorithms \cite{yang}, and PMU-based approaches for transmission-line outage detection and localization \cite{deng}. A related direction of research has focused on forced oscillations, developing methods based on synchrophasor measurements, model-based inference, and physics-informed formulations to identify oscillation sources \cite{huang,lesieutre,frequencyosc,liu}. More recently, hybrid data- and knowledge-driven approaches and machine-learning techniques have been proposed for event detection and localization \cite{boyd,oelhaf,mozaffari,banna,aalam}. 

Despite this considerable progress, there remains a gap as most existing methods are designed only to identify the location of a disturbance without reconstructing its waveform, even though the waveform can provide valuable information about the disturbance's origin, evolution, and physical characteristics. 
Many approaches also rely on detailed physical models or computationally intensive optimization during inference \cite{lesieutre, tyloo}, are tailored to specific classes of events such as forced oscillations or transmission-line outages \cite{deng,huang,lesieutre,frequencyosc,liu,banna}, or are trained and evaluated on predefined disturbance or fault classes \cite{boyd,oelhaf,mozaffari}. Furthermore, most studies consider only single disturbance sources and few demonstrate scalability to large systems while simultaneously handling multiple disturbances under incomplete observation. Table~\ref{tab:method-comparison} summarizes representative disturbance localization methods and qualitatively compares the capabilities demonstrated in the literature, illustrating remaining gaps which serve as the motivation for the current work. 

\begin{table*}[t]
\centering
\begin{threeparttable}
\caption{Comparison of representative power-system disturbance
localization methods. A check mark indicates that
the capability was explicitly demonstrated or claimed in the cited
work; a half-filled circle indicates a partial or restricted capability;
and a dash indicates that the capability was not demonstrated.}
\label{tab:method-comparison}

\scriptsize
\setlength{\tabcolsep}{2.2pt}
\renewcommand{\arraystretch}{1.18}

\begin{tabularx}{\textwidth}{
    >{\raggedright\arraybackslash}m{0.13\textwidth}
    >{\raggedright\arraybackslash}X
    Y Y Y Y Y Y Y
}
\toprule
\textbf{Reference}
&
\textbf{Event type}
&
\makecell{\textbf{Source}\\\textbf{localization}}
&
\makecell{\textbf{Waveform}\\\textbf{reconstruction}}
&
\makecell{\textbf{Physical model-}\\\textbf{free inference}}
&
\makecell{\textbf{Partial}\\\textbf{observations}}
&
\makecell{\textbf{Large-scale}\\\textbf{validation}}
&
\makecell{\textbf{Simultaneous}\\\textbf{disturbances}}
&
\makecell{\textbf{Online}\\\textbf{potential}}
\\
\midrule

You \emph{et al.} \cite{you}
&
General disturbances
&
\yes & \no & \partialyes & \no & \yes & \no & \yes
\\

Lee, Zhang, and Modiano \cite{lee}
&
Power imbalance
&
\yes & \no & \yes & \yes & \no & \partialyes & \no
\\

Yang \emph{et al.} \cite{yang}
&
General disturbances
&
\yes & \no & \yes & \no & \yes & \no & \yes
\\

Deng \emph{et al.} \cite{deng}
&
Line outages
&
\yes & \no & \yes & \no & \no & \no & \yes
\\

Huang \emph{et al.} \cite{huang}
&
Forced oscillations
&
\yes & \no & \yes & \yes & \yes & \no & \partialyes
\\

Lesieutre \emph{et al.} \cite{lesieutre}
&
Forced oscillations
&
\yes & \no & \no & \no & \yes & \no & \partialyes
\\

Delabays \emph{et al.} \cite{frequencyosc}
&
Forced oscillations
&
\yes & \no & \yes & \no & \yes & \yes & \partialyes
\\

Liu, Zhu, and Kekatos \cite{liu}
&
Forced oscillations
&
\yes & \no & \yes & \yes & \yes & \no & \yes
\\

Banna \emph{et al.} \cite{banna}
&
Forced oscillations
&
\yes & \no & \yes & \no & \yes & \no & \yes
\\

Aalam and Shubhanga \cite{aalam}
&
General events
&
\yes & \no & \yes & \no & \partialyes & \no & \yes
\\

Li \emph{et al.} \cite{li}
&
General disturbances
&
\yes & \no & \yes & \no & \yes & \no & \partialyes
\\

\midrule

\textbf{Proposed method}
&
\makecell[l]{\textbf{Time-dependent}\\\textbf{additive power}\\\textbf{disturbances}}
&
\textbf{\yes}
&
\textbf{\yes}
&
\textbf{\yes}
&
\textbf{\yes}
&
\textbf{\yes}
&
\textbf{\yes}
&
\textbf{\partialyes}
\\

\bottomrule
\end{tabularx}

\end{threeparttable}
\vspace{-10pt}
\end{table*}

In this paper, we address the disturbance detection and localization problem using reservoir computing \cite{Echo,rc2}, a recurrent neural-network framework in which only the output layer is trained, enabling efficient training and inference. Reservoir computing has traditionally been used for nonlinear time-series prediction, state estimation \cite{stateEst}, and forecasting of complex dynamical systems \cite{michelle}, and only recently for disturbance detection \cite{restrepo,skardal_restrepo}. Its fast training and inherent memory of recent dynamics make it particularly attractive for real-time monitoring of power-systems dynamics. Here, we use reservoir computing to simultaneously localize and reconstruct the waveforms of multiple disturbance sources in a large-scale power system with incomplete observations, capabilities not jointly demonstrated in prior work, as summarized in the final row of Table~\ref{tab:method-comparison}.

The contributions of this paper are twofold. First, we develop a reservoir-computing framework for power-system disturbance localization that simultaneously identifies multiple concurrent disturbance sources and reconstructs their temporal waveforms from incomplete wide-area measurements. This addresses key limitations of existing disturbance-localization methods. Second, we demonstrate that the proposed framework performs these tasks efficiently using only the grid topology, without requiring an explicit model of its governing system dynamics during inference, enabling fast, data-driven monitoring of large-scale power systems.

\vspace{-0.5em}

\section{Disturbance Localization and identification using Reservoir Computing} \label{sec:rc}

\subsection{Problem Formulation}

The dynamics of a power system can be described in a compact form as:
\begin{equation}
    \frac{d\mathbf{x}}{dt} = \mathbf{F}(\mathbf{x}) + \mathbf{g}(t),
\label{genSystDist}
\end{equation}
where $\mathbf{x}(t)\in\mathbb{R}^{d}$ is the vector of system state variables, $\mathbf{F}\!\colon\!\mathbb{R}^{d}\!\to\!\mathbb{R}^{d}$ describes the unknown undisturbed system dynamics, and $\mathbf{g}(t)\in\mathbb{R}^{d}$ represents an external disturbance.
Even though a real world power grid operates through continuous disturbances, we assume that the disturbances that are sufficiently significant such that their impact reaches a critical threshold are applied only to a small set of nodes, i.e., $g_n(t)\neq 0$ only for $n\in \mathcal{D},$ where $\mathcal{D}$ is the set of disturbed nodes. Typically we will assume $|\mathcal{D}| \leq 3$, corresponding to one to three simultaneous disturbance sources. This provides a controlled setting for evaluating the localization of multiple disturbance sources while allowing the resulting disturbance dynamics to propagate throughout the network. Extending the framework to a larger number of concurrent disturbances is left to future work. Our goal is to localize the source and reconstruct the waveform of the disturbances using only these time-series measurements and a general grid topology of the system. By ``localize the source" we mean to identify the set $\mathcal{D}$ of disturbed nodes, and by ``reconstruct the waveform" we mean to determine its time course, by approximating $g_n(t)$ for all $n\in \mathcal{D}$.

We interpret a disturbance to be any external input that alters the evolution of the system state $\mathbf{x}(t)$. Therefore, we have assumed that disturbances can be modeled as time-dependent, additive functions on the right-hand side of (\ref{genSystDist}), allowing all forms of disturbances to be represented through a common forcing formulation. 

Leveraging the recent developments in the application of reservoir computing in detecting oscillations in complex dynamical networks \cite{restrepo,skardal_restrepo}, the objective in this paper is to solve the inverse problem of localizing and identifying the disturbances given observations of the system variables. In what follows, we describe a pseudo-parallel architecture that allows us to scale our algorithm to larger power grids. 


\subsection{Pseudo-Parallel Reservoirs}
\label{parallel}
As with other machine learning algorithms, reservoir computing needs a training phase. We assume that we can force the system with a \textit{known} forcing function $\mathbf{\hat{g}}(t)$ and collect a time series of the state variables. Namely, the forced system obeys
\begin{equation}
    \frac{d\mathbf{\hat{x}}}{dt} = \mathbf{F}(\mathbf{\hat{x}}) + \mathbf{\hat{g}}(t),
    \label{trainingSystem}
\end{equation}
from which a time series $\{\mathbf{\hat{x}}(-T+i\Delta t)\}_{i=0}^{T/\Delta t}$ is collected, where the training interval is $[-T,0]$ and $1/\Delta t$ is the sampling rate. 

The collected time series is used to force the \textit{reservoir}, a high dimensional, nonlinear dynamical system described by a state vector $\mathbf{r}\!\in\!\mathbb{R}^M.$ Reservoir computing has the advantage that many convenient physical implementations of reservoirs are possible \cite{michelle}. In numerical simulations, the forced reservoir is typically implemented as a recurrent neural network,
\begin{equation}
    \mathbf{r}(t+\Delta t) = h(B\mathbf{r}(t) + W_{\text{in}}\mathbf{H(\mathbf{\hat{x}}}(t))),
    \label{reservior}
\end{equation}
where $h$ is a nonlinear activation function assumed to act component-wise on its input, $B$ is an $M\times M$ matrix that represents the internal structure of the reservoir, $\mathbf{H}\!:\!\mathbb{R}^{2N}\!\to\!\mathbb{R}^{N_\text{in}}$ is an \textit{observation function} that processes the observations into $N_\text{in}$ inputs, and $W_\text{in}$ is an $M\times N_\text{in}$ input matrix. The observation function can represent preprocessing of the observations or account for missing information by removing data from some dynamic elements in the network. For the sake of simplicity, the focus henceforth will be restricted to generators, for which we take the  phase angle $\delta$ and angular frequency $\omega$ as the measured state variables, and for which the disturbance $g_n(t)$ enters only through the frequency equation.

If one uses data from $N$ generators in a system to train a single reservoir, the computational cost of the algorithm increases and its performance decreases as $N$ becomes large. This has led to the proposal to use parallel reservoir architectures when using reservoir computing to predict spatiotemporal chaotic dynamics \cite{michelle}, dynamics on large networked systems \cite{parallelRC}, and to recover disturbances to network-coupled dynamical systems \cite{skardal_restrepo}. In order to ensure that our disturbance localization and identification algorithm is scalable to large power grids, we implement a parallel architecture whereby there are $N$ separate reservoirs labeled $n=1,2,\ldots,N$. Reservoir $n$ is tasked with identifying the disturbance to generator $n$, and it receives its input \textit{only} from generator $n$ and the network neighbors of generator $n$. More precisely, we define the local input set of reservoir $n$ as
\begin{equation}
    \mathcal{N}_n = \{m: K_{nm} > K_0\}\cup\{n\},
    \label{localSet}
\end{equation}
where $K_{nm}$ is the transfer capacity between generators $n$ and $m$, and $K_0$ is a suitably chosen threshold. In our examples, we will choose $K_0$ to be large enough that the local input sets are reasonably small, and in Section~\ref{sec:partialInfo}, we discuss the effect of varying $K_0$.

\begin{figure*}[!htbp]
	\centering
		\includegraphics[width=1.5\columnwidth]{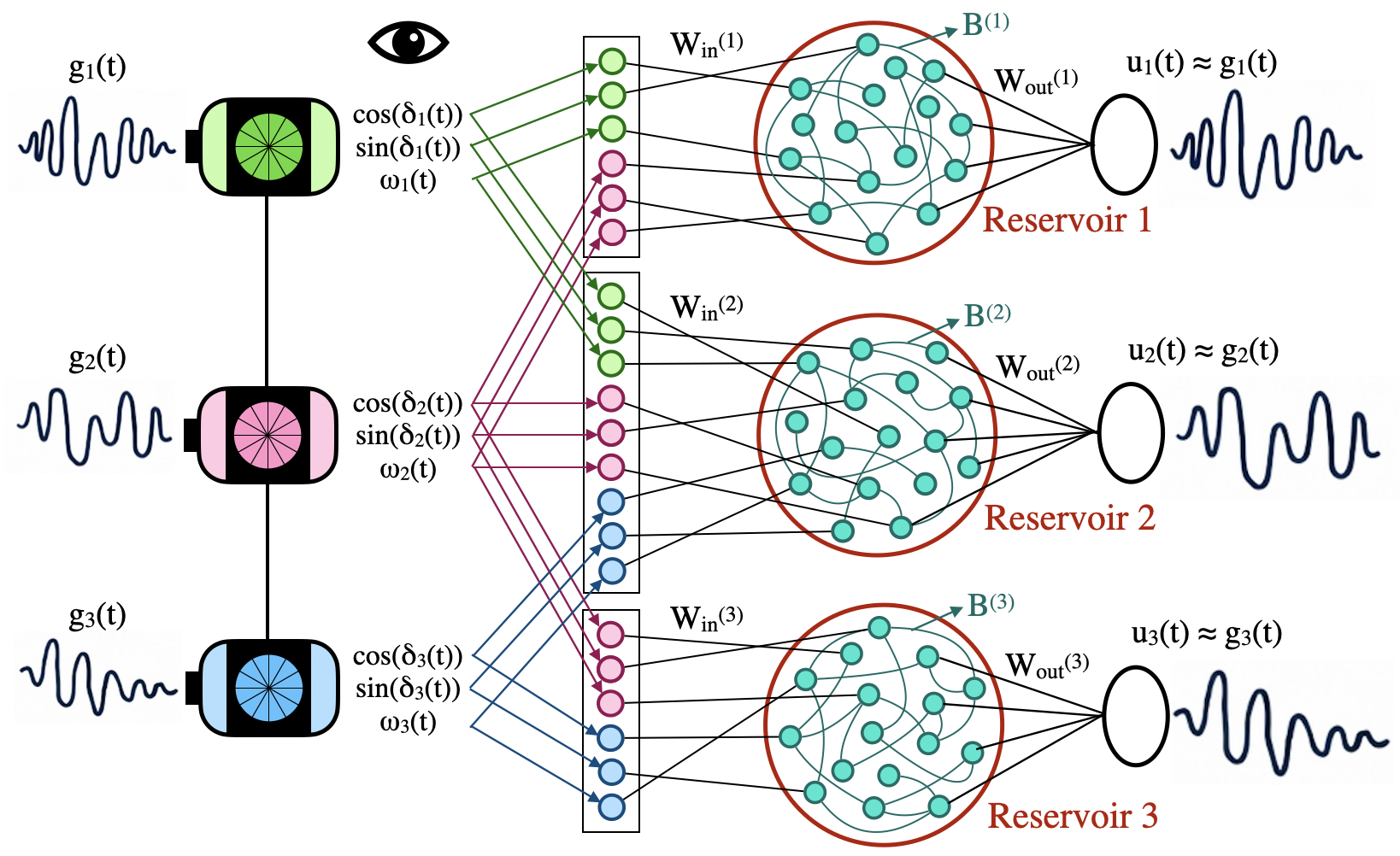} 
	\caption{Schematic illustration of the parallel reservoir computing framework for a system of three coupled synchronous generators. The generators (left) are forced with disturbances (far left). Data observed from each generator is used to force the reservoir of the generators it is connected to. A readout matrix is found for each reservoir to convert its internal state vector into an approximation of the disturbance of its corresponding generator (far right).}
	\label{fig:res}
    \vspace{-10pt}
\end{figure*}

The training of the parallel reservoirs is as follows. Again, collecting a time series $\{\mathbf{\hat{x}}(-T + i\Delta t) \}_{i=0}^{T/\Delta t}$ from (\ref{trainingSystem}), we force the reservoir for generator $n$ as 
\begin{equation}
    \mathbf{r}_n(t+\Delta t) = h(B^{(n)}\mathbf{r}_n(t) + W_\text{in}^{(n)} \mathbf{H}^{(n)}(\mathbf{\hat{x}}(t))),
    \label{input2res}
\end{equation}
where $\mathbf{r}_n\in\mathbb{R}^M$ is the state vector of reservoir $n$, $B^{(n)}$ is the $M\times M$ matrix describing its internal structure, and $W_\text{in}^{(n)}$ is the $M\times N_\text{in}^{(n)}$ input matrix. If we let $\mathcal{N}_n\!=\!\{j_1^{(n)}, j_2^{(n)}, \dots, j_{|\mathcal{N}_n|}^{(n)}\},$ then the observation function
\begin{equation*}
    \mathbf{H}^{(n)}(\mathbf{\hat{x}}) = \begin{pmatrix}
        \mathbf{\hat{x}}_1^{(n)} \\
        \mathbf{\hat{x}}_2^{(n)} \\
        \mathbf{\hat{x}}_3^{(n)}
    \end{pmatrix},
\end{equation*}
concatenates the three vectors
\begin{equation*}
    \begin{aligned}
        \mathbf{\hat{x}}_1^{(n)} &= (\omega_{j_1^{(n)}}, \omega_{j_2^{(n)}},\dots,\omega_{j_{|N_n|}^{(n)}})^T, \\
        \mathbf{\hat{x}}_2^{(n)} &= (\sin\delta_{j_1^{(n)}}, \sin\delta_{j_2^{(n)}},\dots,\sin\delta_{j_{|N_n|}^{(n)}})^T, \\
        \mathbf{\hat{x}}_3^{(n)} &= (\cos\delta_{j_1^{(n)}}, \cos\delta_{j_2^{(n)}},\dots,\cos\delta_{j_{|N_n|}^{(n)}})^T, 
    \end{aligned}
\end{equation*}
that is, the input to reservoir $n$ consists of the frequencies and the sines and cosines of the phase angles of the generators in its local input set. Therefore, $N_\text{in}^{(n)} = 3|\mathcal{N}_n|.$

Once the reservoir has been thus forced with the time series $\{\mathbf{\hat{x}}(-T+i\Delta t)\}_{i=0}^{T/\Delta t}$, a $1\times M$ \textit{output} matrix $W_\text{out}^{(n)}$ is chosen so as to make the output of the reservoir, $u_n(t) = W_\text{out}^{(n)} \mathbf{r}_n(t),$ as close as possible to the known training function $\hat{g}_n(t)$. Note that each reservoir is tasked only with inferring a scalar function, $\hat{g}_n(t).$

To optimize $W_{\text{out}}^{(n)}$ for each reservoir, a ridge regression procedure is used to minimize the cost function
\begin{equation}
    \sum_{i=0}^{T/\Delta t} \|\hat{g}_n(-i\Delta t) - W_\text{out}^{(n)}\textbf{r}_n(-i\Delta t) \|^2 + \lambda \text{Tr}(W_{\text{out}}^{(n)}W_{\text{out}}^{(n)T}),
    \label{ridgeReg}
\end{equation}
where the regularization term $\lambda \geq 0$ prevents overfitting.

Once an output matrix $W_\text{out}^{(n)}$ has been found, an unknown disturbance $\mathbf{g}(t)$ acting on the system as in (\ref{genSystDist}) can be identified by forcing the reservoir with the observed state vector $\mathbf{x}(t)$ as: 
\begin{equation}
\label{eq:reservoir2}
    \mathbf{r}_n(t+\Delta t) = h(B^{(n)}\mathbf{r}_n(t) + W_\text{in}^{(n)} \mathbf{H}^{(n)}(\mathbf{x}(t))).
\end{equation}
Then, using the same $W_{\text{out}}^{(n)}$ found during training, the output 
\begin{equation}
    u_n(t)=W_{\text{out}}^{(n)}\textbf{r}_n(t)
    \label{output}
\end{equation}
is expected to be a good approximation of the disturbance $g_n(t).$ Note that this can be performed online, i.e., the reservoir will produce an approximation to $g_n(t)$ as it is fed with current data. Fig.~\ref{fig:res} shows schematically how the pseudo-parallel reservoir computing method works given a system of three coupled synchronous generators being disturbed by three different forcings.


\subsection{Reservoir Architecture}

Throughout this paper, unless otherwise specified, matrices $B^{(n)}$ and $W_{\text{in}}^{(n)}$ have sparse randomized values, and for simplicity all reservoirs are assumed to have the same size, $M$. The $M\times M$ matrix $B^{(n)}$ is constructed as follows. Entry $B_{ij}^{(n)}$ is chosen to be nonzero with probability $6/M$, and zero otherwise. Each non-zero entry of $B^{(n)}$ is randomly and independently chosen from a uniform distribution in $[-1,1]$. The resulting matrix is then rescaled so that its spectral radius is equal to $\rho = 0.9$. This chosen internal reservoir structure follows that used in \cite{restrepo,skardal_restrepo}. 

Following the work of \cite{michelle}, the input matrix $W_\text{in}^{(n)}\in\mathbb{R}^{M\times N_\text{in}^{(n)}}$ is constructed so that each of the $M$ reservoir nodes receives exactly one of the $N_\text{in}^{(n)}$ inputs. Since the number of reservoir nodes is generally much larger than the number of inputs, each input is connected to multiple reservoir nodes. To control the average number of reservoir nodes assigned to each input, all reservoirs are chosen to have the same size, $$M = S\bar N_\text{in},$$ where $S$ is a scaling parameter and $$\bar N_\text{in} = \frac{1}{N}\sum_{n=1}^N N_\text{in}^{(n)}$$ is the average number of inputs across all reservoirs. 

Within reservoir $n$, the $M$ reservoir nodes are divided as evenly as possible among the $N_\text{in}^{(n)}$ inputs. Therefore, each input is assigned to either $$\left\lfloor \frac{M}{N_\text{in}^{(n)}} \right\rfloor \quad \text{or} \quad \left\lceil \frac{M}{N_\text{in}^{(n)}} \right\rceil$$ reservoir nodes, with any remaining nodes distributed uniformly at random among the inputs. The reservoir nodes assigned to each input are then selected uniformly at random without replacement, ensuring that each reservoir node is connected to exactly one input. Unless otherwise specified, we use $S = 4$. The non-zero elements of $W_{\text{in}}^{(n)}$ are chosen from a uniform random distribution in $[-\sigma,\sigma],$ with $\sigma = 0.1.$ 

The activation function $h$ in (\ref{reservior}) must be a nonlinear function, and it is typically chosen to be a hyperbolic tangent. However, different activation functions work better for different applications \cite{Hurley2023}. In our case, we found numerically that the swish function $$h(\textbf{y}) = \frac{\textbf{y}}{1 + e^{-\eta \textbf{y}}}$$ with $\eta = 0.001$ works better than the hyperbolic tangent. The output matrices $W_\text{out}^{(n)}$ are computed using ridge regression with regularization parameter $\lambda = 10^{-10}.$ Table~$\ref{table:rc-params}$ summarizes all of the parameter values.

\begin{table}[!t]
\centering
\caption{Reservoir computing parameter values.}
\label{table:rc-params}
\begin{tabular}{ccl}
\toprule
Parameter & Value & Description \\
\midrule
$S$        & $4$         & reservoir size scaling parameter \\
$\rho$     & $0.9$       & spectral radius of $B^{(n)}$ \\
$\sigma$   & $0.1$       & input weight distribution parameter \\
$\eta$     & $0.001$     & swish function parameter \\
$\lambda$  & $10^{-10}$  & ridge regression regularization parameter \\
$T$        & $200$       & length of training time interval \\
$\Delta t$ & $0.001$      & time interval step size \\
$K_0$      & $1$         & transfer capacity matrix threshold \\
\bottomrule
\end{tabular}
\vspace{-10pt}
\end{table}

\vspace{-10pt}
\begin{figure*}[!t]
	\centering
		\includegraphics[width=1.85\columnwidth]{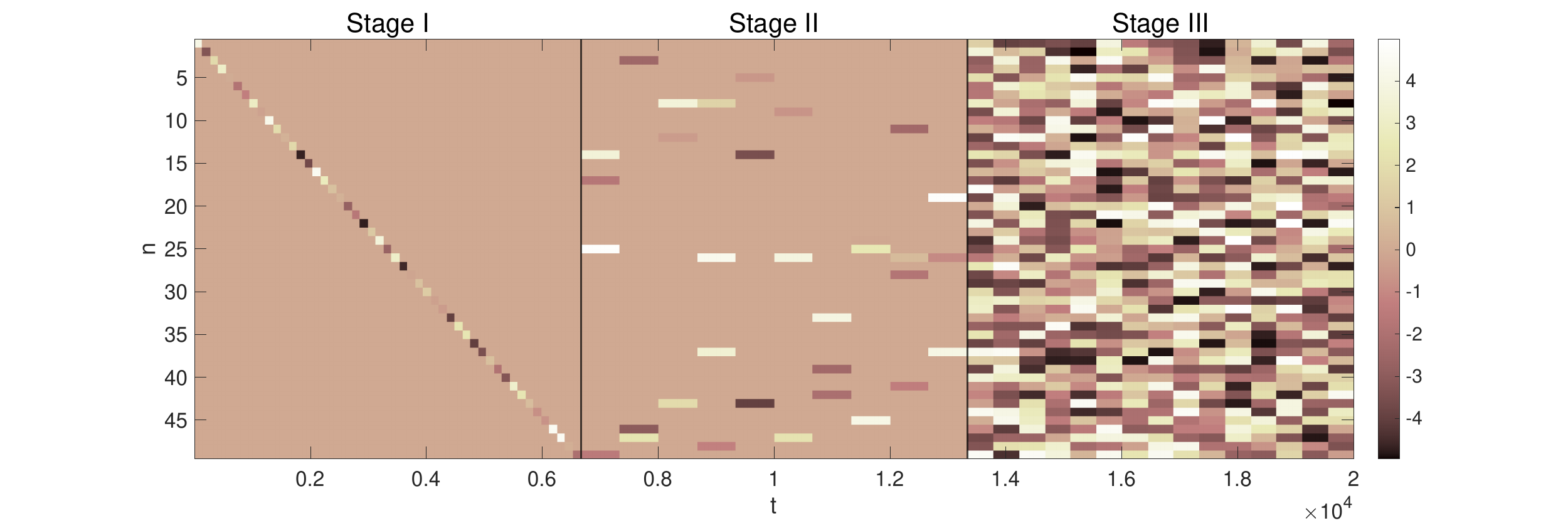}
	\caption{Training forcing function $\hat{g}_n(t)$ as a function of time $t$ (horizontal axis) and generator index $n$ (vertical axis). The training phase is divided into three stages. In the first stage, each generator is forced sequentially with a square pulse of random amplitude. In the second stage, groups of three or four generators are forced simultaneously. In the third stage, all generators are forced simultaneously. See text for more details.}
	\label{fig:wecctraining2}
    \vspace{-10pt}
\end{figure*}


\subsection{Training forcing function} \label{tff}

We choose the training forcing function $\mathbf{\hat{g}}(t)$ to be piecewise constant and organize the training into three stages. In the first stage, $t\in [-T,-2T/3],$ this interval is divided into $N$ equal subintervals. During the $n^\text{th}$ sub-interval, only generator $n$ is forced with a rectangular pulse. Each pulse amplitude $F_n$ is drawn independently from a uniform distribution on the interval $[-5,5]$. More precisely, $$\hat{g}_n(t) = F_n\left[H(t-t_{n-1}) - H(t-t_n)\right],$$ where $$t_n = -T + \frac{nT}{3N},$$
and $H(\cdot)$ is the Heaviside function.

In the second stage, the interval $t\in [-2T/3,-T/3]$ is divided into $L_\text{II}$ equal subintervals, with $L_\text{II} = 50$ in the simulations unless otherwise specified. Within each subinterval, a randomly selected subset of generators is simultaneously forced with a square pulse. Each subset contains either three or four generators, with the subset size chosen independently for each selection with equal probability. The subset of forced generators and pulse amplitudes for each generator are independently resampled for each subinterval, with the amplitudes drawn from a uniform distribution in $[-5,5].$

In the third stage, the interval $t\in [-T/3,0]$ is partitioned into $L_\text{III}$ equal subintervals, with $L_\text{III} = 75$ in the simulations unless otherwise specified. During each subinterval, every generator is forced with a square pulse. The pulse amplitude applied to each generator is independently sampled uniformly from $[-5,5]$, and held constant throughout the subinterval. 

Fig.~\ref{fig:wecctraining2} shows an example of a training forcing function using $L_\text{II}=10$ and $L_\text{III}=15$ with a network of $N=49$ generators for illustration. There are many other potentially good choices for the training forcing function. We chose this specific training protocol so that the reservoirs are trained to situations in which there is only one disturbed generator (stage I), as well as multiple simultaneously disturbed generators (stages II and III). In general, increasing $L_\text{II}$ and $L_\text{III}$ improves recovery by providing the reservoirs with a greater number of distinct training inputs.N


\vspace{-.5em}

\section{Simulation Setup and Implementation} \label{sec:expSet}

\subsection{Test systems}

We tested our algorithm on three different power systems with varying degrees of complexity: 
\begin{itemize}
    \item The IEEE 9-bus system \cite{ieee}
    \item The IEEE 39-bus system \cite{ieee},
    \item The 240-bus reduced Western Electricity Coordinating Council (WECC) model \cite{wecc}.
\end{itemize}

For each system, we obtained parameter values for each generator’s inertia time-constant, damping, and power export parameters, as well as the network’s admittance matrix, and specialize (\ref{trainingSystem}) to the standard swing equations. After Kron reduction and generator aggregation, the 9- and 39-bus IEEE systems and the 240-bus WECC system result in systems with 3, 10, and 49 synchronous machines, respectively. 



\vspace{-1em}

\subsection{Training Procedure}

For each test system, we simulated the corresponding aggregated network using (\ref{trainingSystem}) with the training forcing functions described in Section~\ref{tff} over the interval $t\in[-T,0]$, where $T=200$, using a time step of $\Delta t = 0.001.$ Equation (\ref{trainingSystem}) was solved in MATLAB using Heun's method to generate the training dataset.
The resulting time series was then down-sampled, with only every 10th time step supplied as input to the pseudo-parallel reservoirs described in Section~\ref{parallel}, which evolve according to (\ref{input2res}). Reservoir parameters are listed in Table \ref{table:rc-params}. The readout matrices $W_\text{out}^{(n)}$ were computed using the ridge regression procedure given by (\ref{ridgeReg}), completing the training process.

\begin{table*}[!b]
\centering
\caption{Mean RRMSE and standard error across 200 independent 
realizations, by test system and disturbance type.}
\label{tab:1hz14hz}
\resizebox{\textwidth}{!}{%
\begin{tabular}{lcccccc}
\toprule
& \multicolumn{2}{c}{IEEE 9-bus system} 
& \multicolumn{2}{c}{IEEE 39-bus system} 
& \multicolumn{2}{c}{WECC 240-bus system} \\
\cmidrule(lr){2-3} \cmidrule(lr){4-5} \cmidrule(lr){6-7}
Disturbance type 
& Mean $\left(10^{-1}\right)$ & SE $\left(10^{-3}\right)$
& Mean $\left(10^{-1}\right)$ & SE $\left(10^{-3}\right)$
& Mean $\left(10^{-1}\right)$ & SE $\left(10^{-3}\right)$ \\
\midrule
1~Hz disturbance                    & 0.9 & 26.8 & 1.6 & 6.7 & 0.6 & 0.2 \\
14.7~Hz disturbance             & 11.3 & 121.4 & 3.5 & 10.9 & 8.4 & 1.0 \\
\bottomrule
\end{tabular}%
}
\vspace{-10pt}
\end{table*}


\subsection{Testing Procedure}

After training, the reservoirs were evaluated on a range of previously unseen disturbances. 

\subsubsection{Disturbance scenarios}\label{testDist}

First, two conventional oscillations were applied to all three different test systems:
\begin{itemize}
    \item a 1~Hz disturbance to emulate inter-area oscillations 
    \item a 14.7~Hz disturbance to emulate sub-synchronous oscillations \cite{14hz} 
\end{itemize}    
This was to evaluate the performance of our method and to establish its robustness to different topologies and scalability. Once these features were established, we then narrowed our focus to the third test system (the most complex of the three). We considered three idealized disturbances to test our algorithm, as follows:
\begin{itemize}
    \item a square pulse of random amplitude
    \item a time-localized stochastic disturbance generated from a zero-mean Ornstein--Uhlenbeck process
    \item a biharmonic oscillatory disturbance
\end{itemize}

Finally, we applied three highly-nonlinear disturbances derived from real-world power grid events, as follows:
\begin{itemize}
    \item a data-center load profile digitized from AI training load curves reported by NERC \cite{NERC2025EmergingLoads}
    \item a data-center forced oscillation profile digitized from AI training load curves reported by NERC \cite{NERC2025EmergingLoads} 
    \item an approximately 18~Hz oscillatory disturbance derived directly from measured power system data on July 29, 2021 reported by AEMO for the West Murray Zone \cite{aemo2023westmurray}.
\end{itemize}

It is important to note that, with the exception of the square pulse, the other types of disturbances are different than those used to train the reservoir. As we will see, however, the reservoir successfully identified them.

\subsubsection{Testing methodology}\label{testMethod}

Similarly to training, we simulated the disturbed test system using (\ref{genSystDist}) with the different testing disturbance signals described above. 
For each disturbance configuration, we repeated the full testing procedure over 200 independent realizations, randomly selecting the disturbed generators and other disturbance parameters in each realization, and report statistics averaged across all realizations.

We evaluated the method under two experimental designs, summarized in five cases:
\subsubsection*{\textbf{Design 1} -- same disturbance type across generators (Table \ref{tab:rrmse-by-type})}
\begin{itemize}
    \item Case I: one randomly chosen generator disturbed, repeated for all six disturbance types
    \item Case II: two generators disturbed simultaneously with the same disturbance type, repeated for all six types
    \item Case III: three generators disturbed simultaneouly with the same disturbance type, repeated for all six types
\end{itemize}
\subsubsection*{\textbf{Design 2} -- different disturbance types across generators (Table \ref{tab:rrmse-mixed})}
\begin{itemize}
    \item Case IV: two generators disturbed simultaneously, each assigned a different, randomly chosen disturbance type, both beginning at the same time
    \item Case V: three generators disturbed simultaneously, each assigned a different, randomly chosen disturbance type, all beginning at the same time
\end{itemize}
We tested both experimental designs (Cases I-V) with the WECC system. The results provide a qualitative demonstration of simultaneous disturbance localization and waveform reconstruction followed by a quantitative analysis.

The test data was generated over the interval $t\in[0,T]$, where $T=200$, using a time step of $\Delta t = 0.001.$ Equation (\ref{genSystDist}) was solved in MATLAB using Heun's method. The resulting time series was supplied to the pseudo-parallel reservoirs, which were driven using every 10th simulation time step (corresponding to an effective input interval of $0.01$ time units).
The trained output matrices were held fixed throughout testing and used to approximate the disturbance $g(t)$ for each generator according to (\ref{output}), where $u_n(t)\approx g_n(t)$. 

\vspace{-0.5em}

\section{Results and Discussions} \label{sec:results}


This section presents the results of the proposed reservoir computing framework and discusses the findings in the following order: (A) scalability across power system topologies, (B) localization and reconstruction of nonlinear disturbances, (C) spatial localization accuracy, (D) disturbance reconstruction accuracy, (E) robustness to partial information, (F) robustness to model mismatch, and (G) computational performance. 
Sections (B) through (G) focus exclusively on the reduced WECC 240-bus system, the largest and most realistic of the three test systems under consideration, in order to demonstrate the method's ability to manage complexity in a more realistic power grid. 

As discussed in Section~\ref{intro}, no existing method simultaneously localizes disturbances \emph{and} reconstructs their temporal waveforms (Table~\ref{tab:method-comparison}); therefore, a convenient quantitative comparison against prior methods was not possible, and instead we focused on characterizing the performance of the proposed approach in what follows. 

\vspace{-1em}

\subsection{Scalability across power system topologies}\label{sec:3syst}

In this section, we tested our method on three different systems to show that the proposed method is robust to topology variations and scales up to larger systems. For each disturbed generator $n\in\mathcal{D},$ we calculated the relative root mean square error (RRMSE) between the inferred and actual disturbances, and then averaged it over the disturbed generators to obtain the global RRMSE,
\begin{equation}
    \text{RRMSE} = \frac{1}{|\mathcal{D}|} \sum_{n\in\mathcal{D}} \sqrt{\frac{\left\langle \left[W_\text{out}^{(n)}\mathbf{r}_n(t) - g_n(t)\right]^2\right\rangle_t}{\left\langle g_n(t)^2\right\rangle_t}},
\end{equation}
where $\langle\cdot\rangle_t$ indicates a time average over the time during which the actual disturbance is nonzero. Unless otherwise noted, every RRMSE value reported in this paper is averaged over $200$ realizations, with error bars denoting the standard error of the mean.

We used the RRMSE to quantify waveform reconstruction accuracy because it provides a scale-independent measure of the overall time-domain discrepancy between the true and reconstructed disturbances. As a time-domain metric, however, it was sensitive to timing differences, particularly for higher-frequency disturbances. We therefore used comparisons of power spectral density (PSD) as a complementary measure of frequency-content recovery. 

For this comparison, each realization disturbed the largest generator in the system with both a 1~Hz and a 14.7~Hz disturbance. These results are summarized in Table~\ref{tab:1hz14hz}. The proposed method reconstructed the disturbances on all three systems, with mean RRMSE below $0.16$ for the 1~Hz case and larger values for the 14.7~Hz case. The RRMSE values vary across systems but remain within the same general range, with no clear dependence on system size. In the 9-bus system, a PSD comparison showed that the reservoir recovered the dominant frequency exactly and its peak amplitude to within $17\%$, illustrating that elevated RRMSE can reflect time-domain sensitivity rather than poor frequency-content recovery.


Having demonstrated reconstruction across all three systems, we now focus on the reduced WECC 240-bus system for the remainder of the analysis. Its larger size and greater complexity provide a more challenging and realistic setting in which to examine the scalability of the proposed method.


\subsection{Disturbance localization and waveform reconstruction in the reduced WECC 240-bus System} \label{wecc}



\begin{figure}[h]
\centering
\includegraphics[width=\columnwidth]{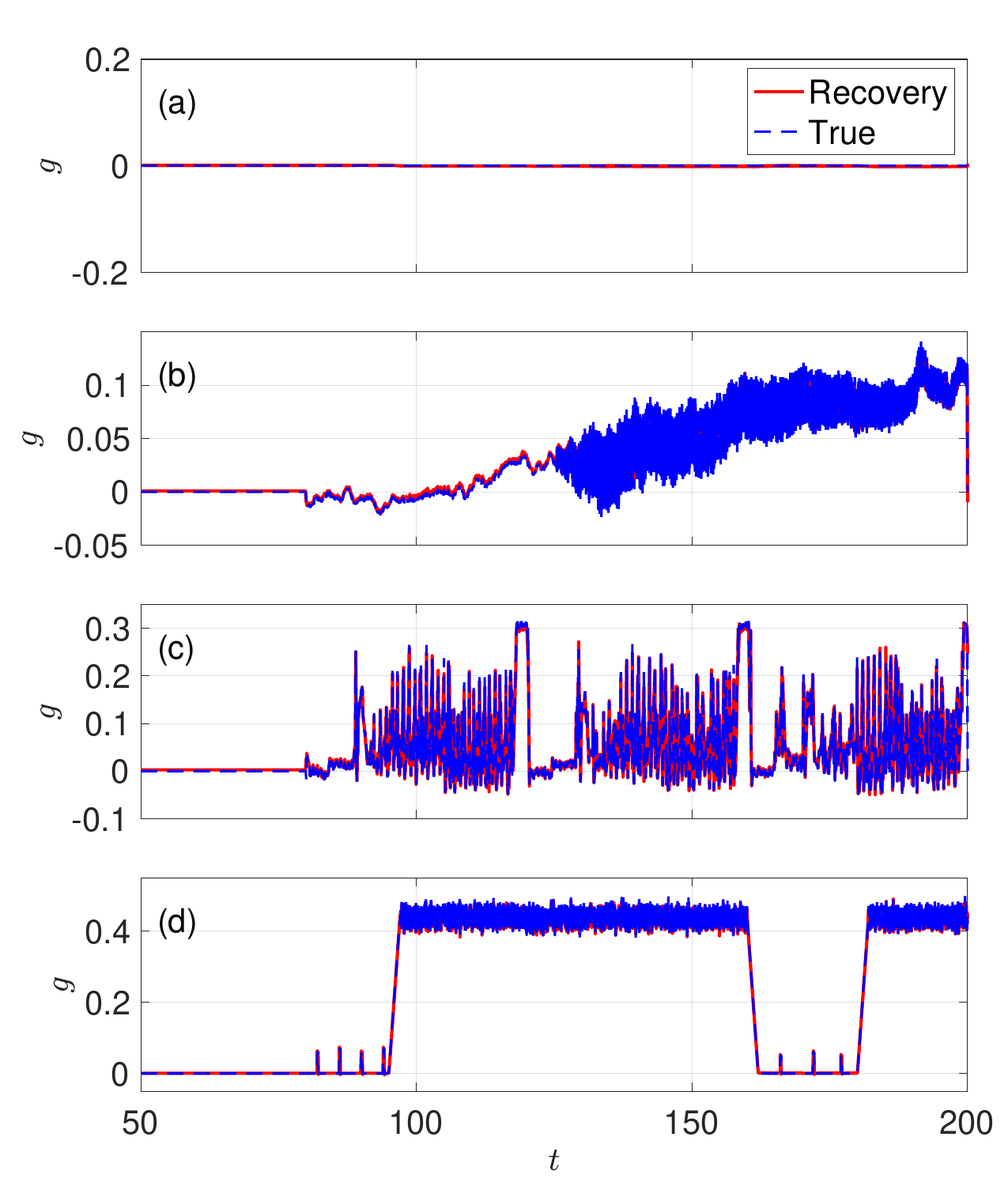}
\caption{Disturbance waveform reconstruction under a mixed-disturbance scenario for the WECC model, in which three generators are disturbed simultaneously by three different disturbance types. The true disturbance $g_n(t)$ (blue dotted lines) and the reservoir's inferred disturbance $W_\text{out}^{(n)}\mathbf{r}_n(t)$ (red solid lines) are shown for an undisturbed generator (a) and for the three disturbed generators (b)--(d): generator (b) is subjected to an AEMO-derived oscillatory disturbance, generator (c) is subjected to a NERC-derived periodic forced oscillation disturbance, and generator (d) is subjected to a NERC-derived AI-training ramp disturbance. The reservoir correctly localizes and reconstructs each disturbance despite their differing type.}
\label{fig:weccvaryG}
\end{figure}

In this section, we evaluate the proposed method using real-world, nonlinear disturbance signals in the reduced WECC 240-bus system. As a representative example, we considered Case V, in which three generators are disturbed simultaneously, each with a different real-world disturbance type described in Section~\ref{testDist}: generator (b) is disturbed by the AEMO oscillatory disturbance, generator (c) by the NERC forced-oscillation disturbance, and generator (d) by the NERC data-center load-ramp disturbance. All five cases (I-V) described in Section~\ref{testMethod} will be tested and discussed later in this section. All three disturbances began simultaneously at $t_0 = 80$~s. Fig.~\ref{fig:weccvaryG} shows the true disturbance $g_n(t)$ (blue dotted lines) alongside the disturbance identified by the reservoir (red solid lines) for the three disturbed generators and one undisturbed generator. Despite the disturbed generators experiencing disturbances of distinct type, frequency content, and amplitude at the same time, the reservoir accurately reconstructs each one individually in panels (b)--(d), while correctly identifying the absence of a disturbance in panel (a); the same result is obtained for all other undisturbed generators - hence not shown. 

\begin{figure}[h]
\centering
\includegraphics[width=1.05\columnwidth]{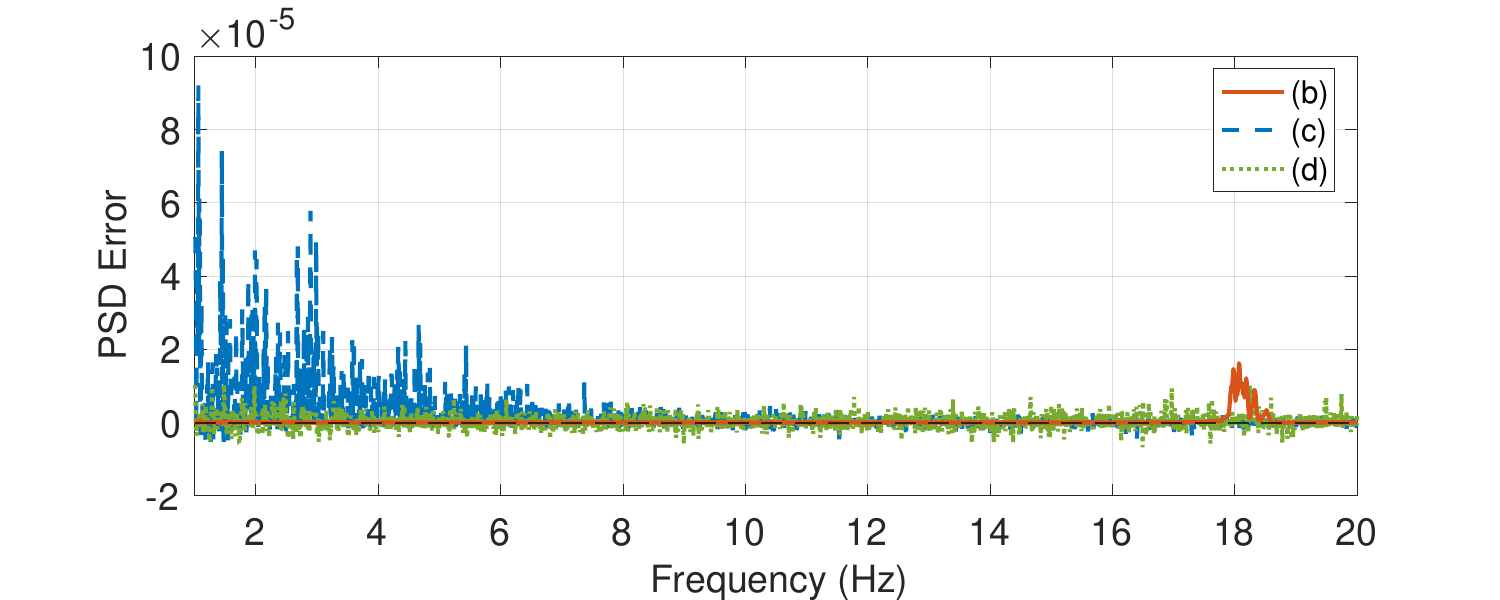}
\caption{Power spectral density (PSD) error for three disturbed generators, calculated as the difference between the true and recovered PSDs at each frequency. The curves correspond to panels (b)--(d) in Fig.~\ref{fig:weccvaryG}.}
\vspace{-10pt}
\label{fig:weccPSD}
\end{figure}

The corresponding RRMSEs for generators (b)--(d) are 0.18, 0.13, and 0.03, respectively. Fig.~\ref{fig:weccPSD} shows the corresponding PSD errors, which remain below $1\times 10^{-4}$ across the frequencies examined, confirming accurate recovery of the dominant frequency content up to 20~Hz. We note that although higher-frequency components were present, their amplitudes were substantially smaller, hence negligible. Therefore, we restricted our attention to the frequency range up to 20~Hz. 


\vspace{-10pt}



\subsection{Spatial localization accuracy}

\begin{figure}[h]
	\centering
    \includegraphics[width=1.05\columnwidth]{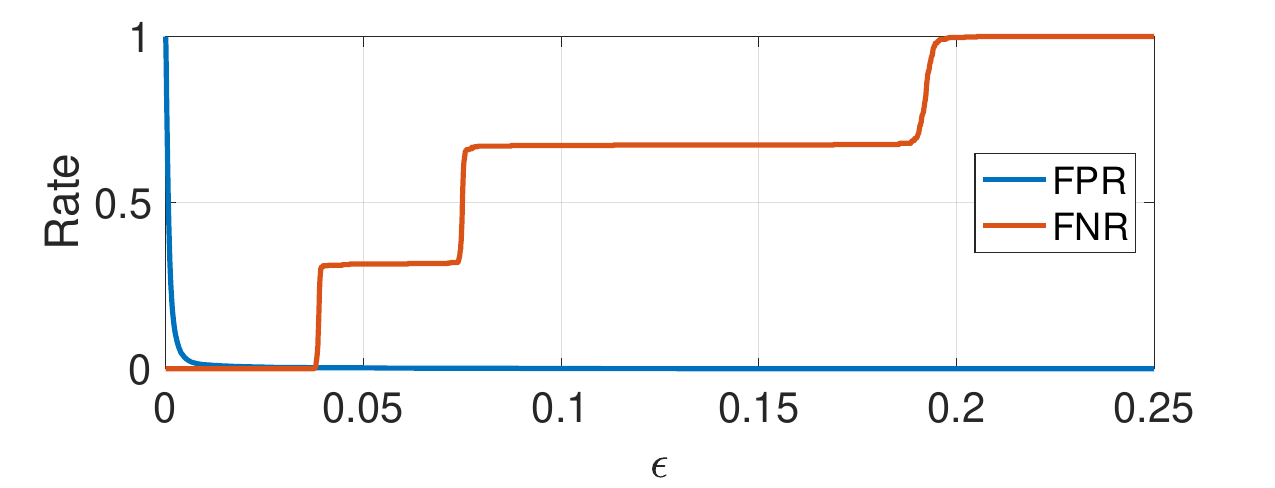}
	\caption{False Positive Rate (blue line) and False Negative Rate (red line) as a function of the tolerance $\epsilon$. Note that since we only disturbed three generators, the False Negative Rate must be a multiple of $1/3$.}
	\label{fig:roc}
\end{figure}

Now we turn our attention to accuracy of localization for different tolerance values. To measure the accuracy of the disturbance localization, we set a tolerance $\epsilon$, and declared generator $n$ as disturbed if the standard deviation of the inferred disturbance was larger than $\epsilon$, i.e., if
$$\frac{\Delta t}{200}\sum_{i=0}^{200/\Delta t} \left[W_\text{out}^{(n)}\mathbf{r}_n(i\Delta t)\right]^2 > \epsilon^2,$$ 
and undisturbed otherwise. We use the standard deviation of the inferred disturbance as a simple measure of the magnitude of the recovered signal to differentiate between disturbed and undisturbed generators. We then computed the number of False Positives (generators falsely classified as disturbed) and False Negatives (generators falsely classified as undisturbed). Finally, we calculated the False Positive Rate and the False Negative Rate by dividing these counts by $N-|\mathcal{D}|=46$ and $|\mathcal{D}| =3$, respectively, and plotted both as a function of the threshold $\epsilon$. The results are shown in Fig.~\ref{fig:roc}, where each realization disturbs three generators with the real-world disturbances described in Section~\ref{testDist}. 

The results indicate that the False Positive Rate approaches $1$ as $\epsilon\rightarrow 0^+,$ while the False Negative Rate approaches $1$ as $\epsilon\rightarrow \infty$ (in our case, for $\epsilon \gtrsim 1$). However, there is a range of values of $\epsilon$ where both rates are zero, demonstrating that robust localization of the disturbances can be achieved by the reservoir. This confirms quantitatively the distinction seen in Fig.~\ref{fig:weccvaryG} between the undisturbed generator (a) and the disturbed generators (b)--(d). Although there is a small range of $\epsilon$ values for which the classification is perfect, plotting the False Positive and False Negative Rates as a function of $\epsilon$ provides a more direct view of the method's performance and sensitivity to the choice of tolerance. The corresponding ROC curve -- which maps the true positive rate against the false positive rate -- has an area under the curve (AUC) of $0.9981$.


\subsection{Disturbance reconstruction accuracy}


Having demonstrated reconstruction qualitatively in Fig.~\ref{fig:weccvaryG}, we now quantify reconstruction accuracy using the RRMSE metric across the two experimental designs and all five cases introduced in Section~\ref{testMethod}. Table~\ref{tab:rrmse-by-type} reports the mean RRMSE and standard error under Design I (Cases I-III), in which each disturbed generator is subjected to the same disturbance type. Table~\ref{tab:rrmse-mixed} reports the corresponding results under Design II (Cases IV-V), in which each disturbed generator is subjected to a different disturbance type.

\begin{table*}[!t]
\centering
\caption{Mean RRMSE and standard error for Design I, varying the number of simultaneously disturbed generators and disturbance type, as described in 
Section~\ref{testMethod}.}
\label{tab:rrmse-by-type}
\resizebox{\textwidth}{!}{%
\begin{tabular}{lcccccc}
\toprule
& \multicolumn{2}{c}{Case I} 
& \multicolumn{2}{c}{Case II} 
& \multicolumn{2}{c}{Case III} \\
\cmidrule(lr){2-3} \cmidrule(lr){4-5} \cmidrule(lr){6-7}
Disturbance type 
& Mean $\left(10^{-1}\right)$ & SE $\left(10^{-3}\right)$
& Mean $\left(10^{-1}\right)$ & SE $\left(10^{-3}\right)$
& Mean $\left(10^{-1}\right)$ & SE $\left(10^{-3}\right)$ \\
\midrule
Square pulse                    & 0.4 & 3.3 & 0.4 & 4.8 & 0.5 & 11.5 \\
Ornstein--Uhlenbeck           & 2.0 & 10.5 & 2.1 & 18.3 & 2.1 & 13.8 \\
Biharmonic oscillatory          & 0.3 & 2.9 & 0.3 & 2.4 & 0.4 & 3.1 \\
NERC data-center demand profile & 0.4 & 0.6 & 0.5 & 2.4 & 0.5 & 6.3 \\
NERC forced oscillation         & 1.4 & 0.8 & 1.9 & 39.8 & 2.1 & 46.4 \\
AEMO oscillatory                & 1.9 & 1.5 & 2.0 & 1.2 & 2.0 & 1.0 \\
\bottomrule
\end{tabular}%
}
\vspace{-5pt}
\end{table*}

\begin{table}[h]
\centering
\caption{Mean RRMSE and standard error for Design II, with different disturbance types applied to each disturbed generator, as described in Section~\ref{testMethod}.}
\label{tab:rrmse-mixed}
\begin{tabular}{lcc}
\toprule
Scenario & Mean $\left(10^{-1}\right)$ & SE $\left(10^{-3}\right)$ \\
\midrule
Case IV & 1.3 & 3.4 \\
Case V & 1.3 & 2.7 \\
\bottomrule
\end{tabular}
\vspace{-10pt}
\end{table}

RRMSE varies by disturbance type, with the Ornstein-Uhlenbeck, NERC forced oscillation, and AEMO oscillatory disturbances producing higher values (0.14-0.21) than the square pulse, biharmonic oscillatory, and NERC load profile disturbances (0.03-0.05). The higher-RRMSE disturbances contain faster, higher-frequency, or irregular fluctuations, making them more sensitive to phase and timing differences. As discussed above, the larger RRMSE values do not necessarily indicate poor recovery; the PSD error in Fig.~\ref{fig:weccPSD} remains below $1\times 10^{-4}$ for the NERC and AEMO disturbances, indicating accurate recovery of their dominant frequency content.

\begin{figure}[h]
	\centering
		\includegraphics[width=1.05\columnwidth]{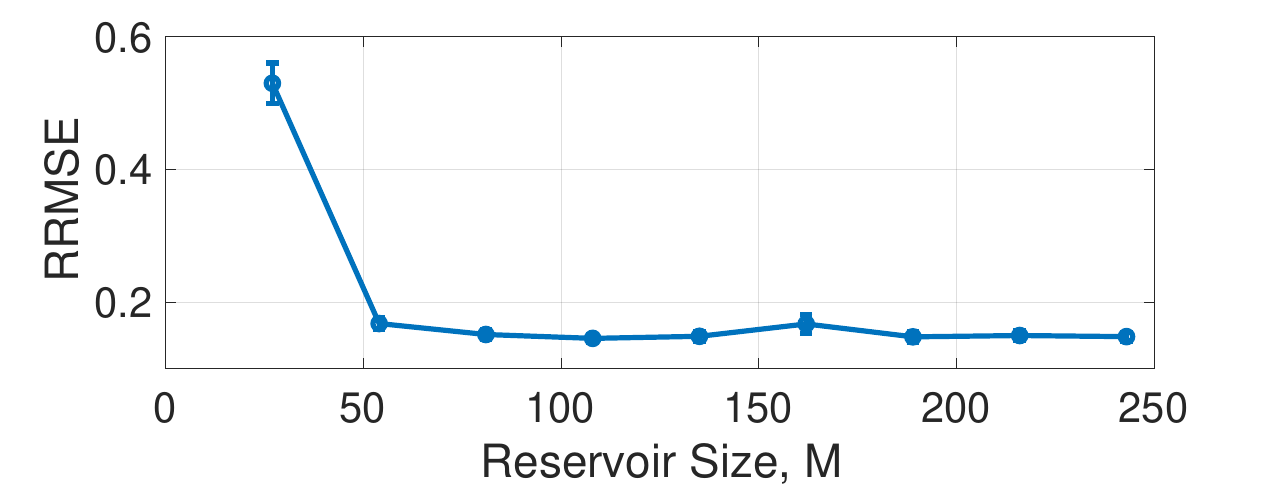}
	\caption{Relative root mean squared error between the inferred and true disturbances as a function of reservoir size, $M$.}
	\label{fig:err2}
    \vspace{-5pt}
\end{figure}

We next examined how reconstruction accuracy depends on the reservoir size $M.$ Fig.~\ref{fig:err2} plots the average RRMSE as a function of the reservoir size $M$, which was varied by adjusting the scaling parameter $S$, over 200 realizations. The results suggest that RRMSE decreases sharply with increasing $M$, falling below 0.2 for $M\gtrsim 50$, corresponding to $S=2,$ with limited improvement for larger reservoirs. This indicates that moderate-sized reservoirs are sufficient for accurate disturbance recovery, highlighting the computational efficiency of the pseudo-parallel architecture, which avoids the need for a single large reservoir to achieve similar accuracy. These results demonstrate successful disturbance localization and waveform reconstruction under ideal operating conditions. The following section evaluates robustness under more realistic scenarios such as missing information and model mismatch. 

\subsection{Robustness to partial information}
\label{sec:partialInfo}

Power grids are comprised of thousands of components, often spread across large geographical footprints, 
making system-wide real-time measurements susceptible to failures, delays, and inaccuracies. In light of this limitation, we investigate the robustness of the proposed method to partial information in two ways: by masking measurements from a subset of generators and by varying the threshold of the transfer capacity matrix used to define the local input sets [see (\ref{localSet})].

\begin{figure}[h]
   \centering

   \begin{subfigure}[t]{0.5\textwidth}
       \centering
       \includegraphics[width=1.05\textwidth]{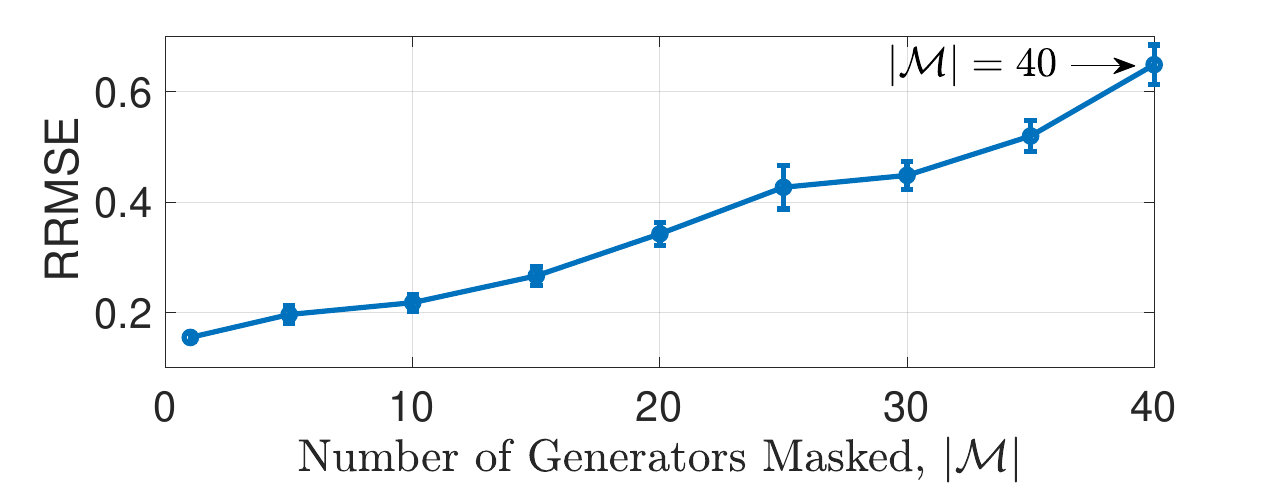}
       \label{fig:first11}
   \end{subfigure}
   \begin{subfigure}[t]{0.5\textwidth}
       \centering
       \vspace{-10pt}
       \includegraphics[width=1.05\textwidth]{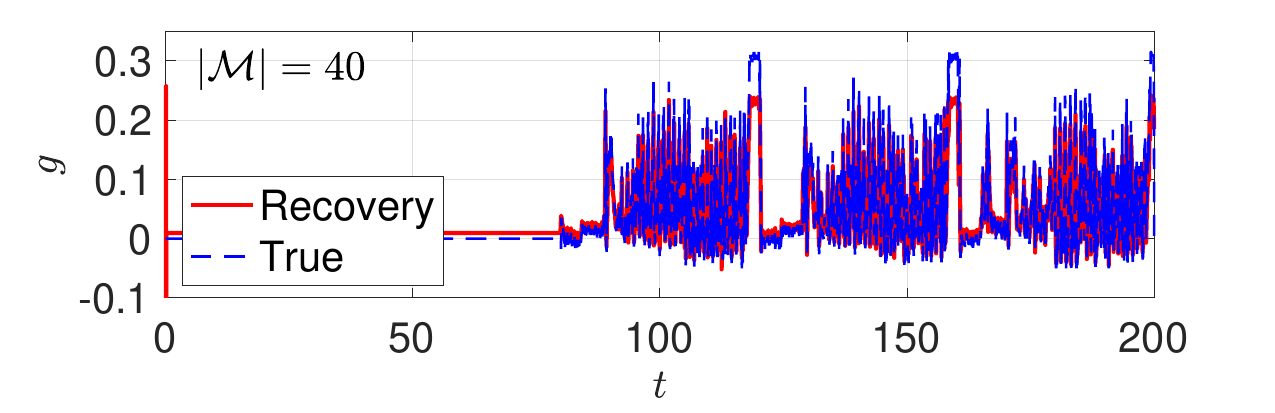}
       \label{fig:second22}
   \end{subfigure} 
	\caption{Relative root mean squared error between the inferred and true disturbances as a function of the number of masked generators, $|\mathcal{M}|$. The bottom panel shows recovery from a disturbed generator when information from only $9$ out of the $49$ generators is available.}
	\label{fig:mask}
\end{figure}

For our first method of partial information testing, we simulated the full system using \textit{all} generators, but provided the reservoir algorithm data from only a subset of them. If the system has $N$ generators, we would normally create $N$ reservoirs, one for each generator, and feed each reservoir data from its local input set. Instead, we masked the data from a subset $\mathcal{M}$ of generators, i.e., we considered information from these generators to be inaccessible. Consequently, we aimed to identify and localize only disturbances to the unmasked generators, $n\notin \mathcal{M}.$ Recovery of disturbances applied to masked generators remains future work. We associated reservoirs to each one of these generators, but redefined the local input set of reservoir $n$ as
$$\tilde{\mathcal{N}}_n = \mathcal{N}\setminus\mathcal{M}, \quad n\notin \mathcal{M},$$
i.e., we did not include information from the masked generators when training the reservoirs. 

Fig.~\ref{fig:mask} shows the RRMSE as a function of the number of masked generators, $|\mathcal{M}|$. Although the RRMSE increases with $|\mathcal{M}|$, the error remains low even when measurements from most generators are unavailable. With measurements from only $9$ out of the $49$ generators available ($|\mathcal{M}| = 40$), the inferred disturbance remains close to the true disturbance (bottom panel). This robustness to incomplete measurements is important for practical deployment, where measurements are often unavailable from all parts of the system.

\begin{figure}[h]
	\centering
		\includegraphics[width=1.05\columnwidth]{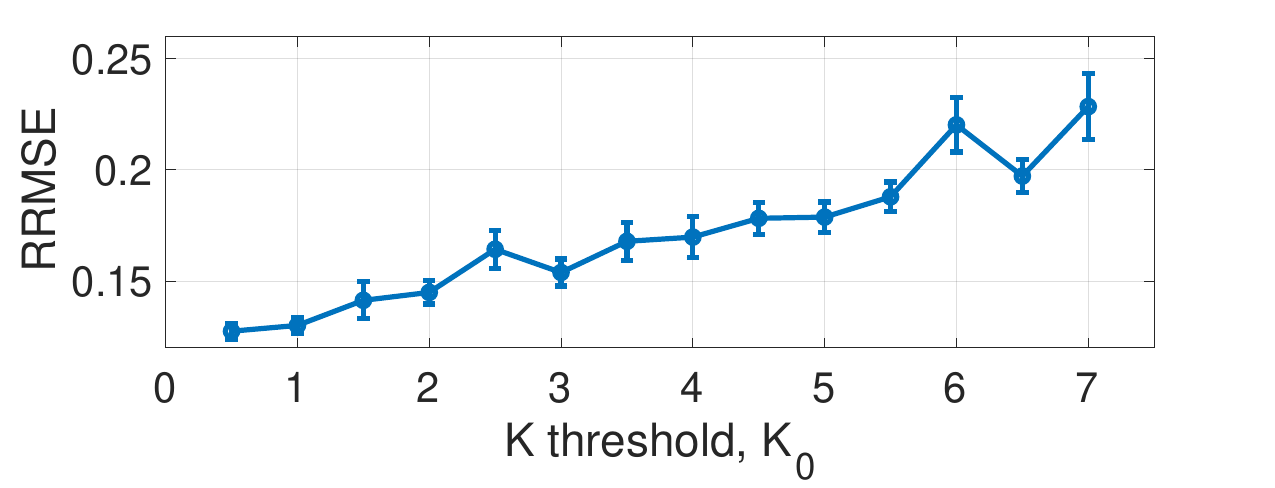}
	\caption{Relative root mean squared error between the inferred and true disturbances as a function of the threshold on the transfer capacity matrix $K_0$ which determines the local input set for each reservoir, $\mathcal{N}_n$.}
	\label{fig:KthreshRMSE}
    \vspace{-5pt}
\end{figure}

Recall that the local input set of reservoir $n,$ $\mathcal{N}_n$ is defined using the transfer-capacity threshold $K_0$ in (\ref{localSet}). We varied $K_0$ to restrict each reservoir to information from increasingly stronger connections, reducing the maximum number of connected generators from 23 at the lowest threshold to 9 at the highest. Fig.~\ref{fig:KthreshRMSE} shows the RRMSE as a function of $K_0.$ Although performance deteriorates as $K_0$ increases, successful reconstruction is maintained with an RRMSE below 0.25. Using fewer connections also reduces the computational cost and suggests that the method can perform well without requiring the exact grid structure.

\vspace{-5pt}

\subsection{Robustness to model mismatch}
\label{noiseResults}

So far we have assumed that the system used to train the reservoir is the same as the one for which the disturbances are later identified. In practice, however, the model used for training may not exactly match the true system. For example, one might train the reservoir using a calibrated model whose parameters differ from those of the physical power grid, or the system parameters may change over time even if the reservoirs are trained using the actual system. It is therefore important to assess the robustness of the proposed method to parameter mismatch between training and testing. 

To address this, we first trained the reservoirs for the WECC system using the nominal generator inertia $H_n$ and damping coefficient $D_n$. We then recovered disturbance waveforms using perturbed inertia and damping parameters,
\begin{equation*}
    \begin{aligned}
        H_n^{\text{test}} &= \max(H_n(1+\nu r_n),0), \\
        D_n^{\text{test}} &= \max (D_n(1+\nu s_n),0),
    \end{aligned}
\end{equation*}
where $r_n$ and $s_n$ are independent random variables drawn from a standard Gaussian distribution, and $\nu$ controls the magnitude of parameter mismatch.

\begin{figure}[h]
	\centering
		\includegraphics[width=1.05\columnwidth]{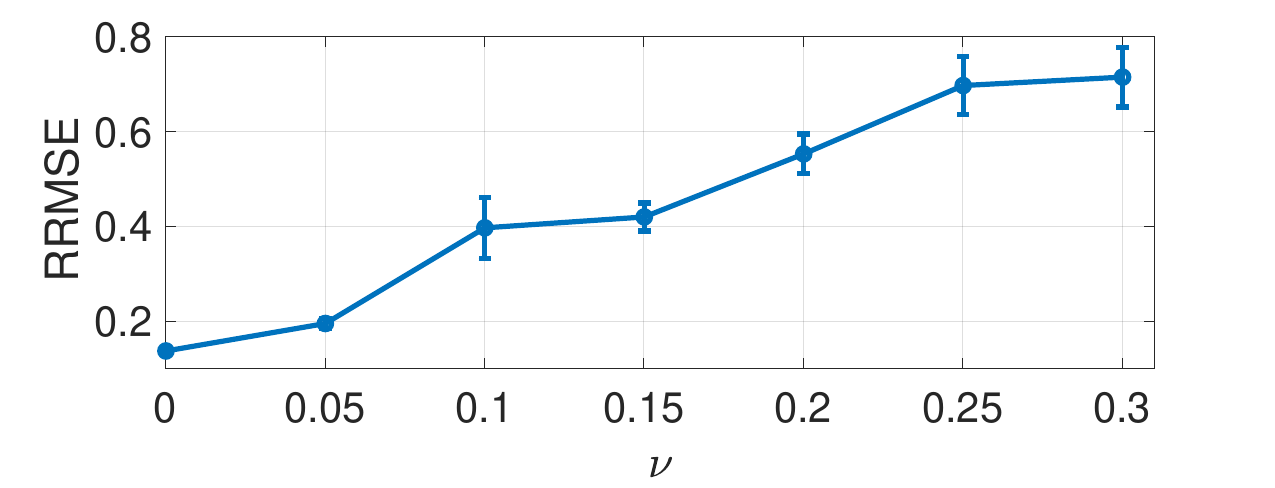}
	\caption{Relative root mean squared error of recovery as the magnitude of mismatch $\nu$ added to inertia and damping parameters increases.}
	\label{fig:rmse}
    \vspace{-10pt}
\end{figure}

Fig.~\ref{fig:rmse} presents the RRMSE as a function of $\nu$. As the mismatch between the training and testing parameters increases, the RRMSE increases; however, the method remains accurate for moderate parameter variations. For example, a $5\%$ mismatch ($\nu = 0.05$) has little effect on the reconstruction error, while even a $10\%$ mismatch maintains a low RRMSE. These results demonstrate that the proposed method is robust to moderate uncertainty in system parameters, supporting its applicability when the model used for training does not exactly match the physical system.

\vspace{-5pt}

\subsection{Computational Performance}
We recorded the training and testing times for each experiment. All computations were performed in MATLAB R2024a on a 2023 14-inch MacBook Pro equipped with an Apple M2 Pro processor and 16 GB of unified memory running macOS Sonoma 14.8.3. Averaged over all experiments, the mean training time was 25.5 seconds, while the mean testing time was 128.5 seconds. These runtimes support the practical use of the proposed approach for online monitoring, where rapid training and inference are essential for timely disturbance detection.

\vspace{-1em}
\section{Conclusion} \label{sec:discussion}
In this work, we utilized reservoir computing to localize the source and reconstruct the waveform of disturbances impacting power systems. Validation was performed using standard IEEE test systems and the reduced 240-Bus WECC system. Additional numerical experiments indicate that the proposed approach scales effectively to networks containing more than 100 synchronous generators, though that is left for future work. The method can not only identify which generators are disturbed, but also recover the dynamics of those disturbances, highlighting its effectiveness in categorizing different disturbance waveforms.
%
More broadly, the framework is not restricted to generator phase and frequency measurements and could be naturally extended to systems with inverter-based resources, where disturbance identification and mitigation are increasingly critical. 




\section*{Acknowledgment}

The authors would like to thank Marena Trujillo, Fiona Majeau, and Matt Baughman, as well as the APPM department at CU Boulder.
This work was authored in part by the National Laboratory of the Rockies for the U.S. Department of Energy (DOE) under Contract No. DE-AC$36$-$08$GO$28308$. The views expressed in the article do not necessarily represent the views of the DOE or the U.S. Government. The U.S. Government retains and the publisher, by accepting the article for publication, acknowledges that the U.S. Government retains a nonexclusive, paid-up, irrevocable, worldwide license to publish or reproduce the published form of this work, or allow others to do so, for U.S. Government purposes.

\ifCLASSOPTIONcaptionsoff
  \newpage
\fi

\vspace{-1em}
\bibliographystyle{IEEEtran}
\bibliography{references}

@online{texasFreeze2021,
  author  = {Andrew Weber and Mose Buchele},
  title   = {Texas has an official death count from the 2021 blackout. The true toll may never be known.},
  journal = {Texas Standard},
  year    = {2022},
  month   = {Aug},
  day     = {5},
  url     = {https://texasstandard.org/stories/texas-freeze-winter-storm-2021-death-count/},
  urldate = {2026-08-03},
}

@techreport{spainBlackout,
  author      = {{ICS Investigation Expert Panel}},
  title       = {Factual Report: Grid Incident in Spain and Portugal on 28 April 2025},
  institution = {Independent Commission on System Events},
  type        = {Technical Report},
  year        = {2025}
}

@article{you,
author = {Shutang You  and Yong Liu  and Micah J. Till  and Jiecheng Zhao  and Wenxuan Yao  and Dao Zhou  and Ling Wu  and Yu Su  and Xuemeng Zhang  and Yilu Liu },
title = {Disturbance location determination based on electromechanical wave propagation in FNET/GridEye: a distribution-level wide-area measurement system},
journal = {IET Generation, Transmission \& Distribution},
volume = {11},
issue = {18},
pages = {4436-4443},
year = {2017},
doi = {10.1049/iet-gtd.2016.1851}
}

@inproceedings{lee,
  author    = {Hyang-Won Lee and Jianan Zhang and Eytan Modiano},
  title     = {Data-Driven Localization and Estimation of Disturbance in the Interconnected Power System},
  booktitle = {2018 IEEE International Conference on Communications, Control, and Computing Technologies for Smart Grids (SmartGridComm)},
  pages      = {1--6},
  year       = {2018},
  doi        = {10.1109/SmartGridComm.2018.8587509}
}

@article{yang,
  title={Data‐driven online distributed disturbance location for large‐scale power grids},
  author={Zekun Yang and Yu Chen and Ning Zhou and Aleksey S. Polunchenko and Yilu Liu},
  journal={IET Smart Grid},
  year={2019},
  doi={10.1049/iet-stg.2018.0081}
}

@inproceedings{deng,
author = {Deng, Xianda and Bian, Desong and Yao, Wenxuan and Jiang, Zhihao and Liu, Yilu},
year = {2019},
month = {05},
pages = {3373-3378},
title     = {Line Outage Detection and Localization via Synchrophasor Measurement},
booktitle = {Proceedings of the IEEE PES Innovative Smart Grid Technologies Asia (ISGT Asia)},
doi = {10.1109/ISGT-Asia.2019.8881142}
}

@article{huang,
  author  = {Tong Huang and Nikolaos M. Freris and P. R. Kumar and Le Xie},
  title   = {A Synchrophasor Data-Driven Method for Forced Oscillation Localization Under Resonance Conditions},
  journal = {IEEE Transactions on Power Systems},
  volume  = {35},
  number  = {5},
  pages   = {3927--3939},
  year    = {2020}
}

@inproceedings{lesieutre,
  author    = {B. C. Lesieutre},
  title     = {Model-Enhanced Localization of Forced Oscillation Using PMU Data},
  booktitle = {Proceedings of the 58th Annual Allerton Conference on Communication, Control, and Computing},
  year      = {2022}
}

@article{frequencyosc,
  title = {Locating the Source of Forced Oscillations in Transmission Power Grids},
  author = {Delabays, Robin and Lokhov, Andrey Y. and Tyloo, Melvyn and Vuffray, Marc},
  journal = {PRX Energy},
  volume = {2},
  issue = {2},
  pages = {023009},
  numpages = {9},
  year = {2023},
  month = {Jun},
  publisher = {American Physical Society},
  doi = {10.1103/PRXEnergy.2.023009},
  url = {https://link.aps.org/doi/10.1103/PRXEnergy.2.023009}
}

@article{liu,
author = {Liu, Shaohui and Zhu, Hao and Kekatos, Vassilis},
year = {2024},
month = {06},
title = {Data-driven forced oscillation localization using inferred impulse responses},
volume = {234},
journal = {Electric Power Systems Research},
doi = {10.1016/j.epsr.2024.110759}
}

@article{boyd,
  author  = {Jonathan D. Boyd and Joshua H. Tyler and Anthony M. Murphy and Donald R. Reising},
  title   = {Learning from Power Signals: An Automated Approach to Electrical Disturbance Identification within a Power Transmission System},
  journal = {Sensors},
  volume  = {24},
  number  = {2},
  pages   = {483},
  year    = {2024},
  doi     = {10.3390/s24020483}
}

@article{oelhaf,
title = {A scoping review of machine learning applications in power system protection and disturbance management},
journal = {International Journal of Electrical Power \& Energy Systems},
volume = {172},
pages = {111257},
year = {2025},
doi = {https://doi.org/10.1016/j.ijepes.2025.111257},
author = {Julian Oelhaf and Georg Kordowich and Mehran Pashaei and Christian Bergler and Andreas Maier and Johann Jäger and Siming Bayer}
}

@Article{mozaffari,
AUTHOR = {Mozaffari, Mahsa and Doshi, Keval and Yilmaz, Yasin},
TITLE = {Real-Time Detection and Classification of Power Quality Disturbances},
JOURNAL = {Sensors},
VOLUME = {22},
YEAR = {2022},
NUMBER = {20},
ARTICLE-NUMBER = {7958},
DOI = {10.3390/s22207958}
}

@article{banna,
author = {Ul Banna, Hasan and Solanki, Sarika Khushalani and Solanki, Jignesh},
title = {Data-driven disturbance source identification for power system oscillations using credibility search ensemble learning},
journal = {IET Smart Grid},
volume = {2},
number = {2},
pages = {293-300},
doi = {https://doi.org/10.1049/iet-stg.2018.0092},
year = {2019}
}

@article{aalam,
title = {Power system event detection and localization—A new approach},
journal = {Electric Power Systems Research},
volume = {223},
pages = {109553},
year = {2023},
doi = {https://doi.org/10.1016/j.epsr.2023.109553},
author = {Mir Khadim Aalam and K.N. Shubhanga}
}

@ARTICLE{tyloo,
  author={Tyloo, Melvyn and Vuffray, Marc and Lokhov, Andrey Y.},
  journal={IEEE Transactions on Power Systems}, 
  title={Forced Oscillation Source Localization From Generator Measurements}, 
  year={2026},
  volume={41},
  number={2},
  pages={1473-1483},
  doi={10.1109/TPWRS.2025.3615451}}

@article{li,
author = {Li, Zikang and Tian, Jiyang and Liu, Hao},
title = {A joint data and knowledge-driven method for power system disturbance localisation},
journal = {IET Generation, Transmission \& Distribution},
volume = {18},
number = {24},
pages = {4078-4089},
doi = {https://doi.org/10.1049/gtd2.13331},
year = {2024}
}

@article{restrepo,
  author  = {J. G. Restrepo and C. P. Byers and P. S. Skardal},
  title   = {Suppressing Unknown Disturbances to Dynamical Systems Using Machine Learning},
  journal = {Communications Physics},
  volume  = {7},
  pages   = {415},
  year    = {2024}
}

@article{skardal_restrepo,
  author  = {Per Sebastian Skardal and Juan G. Restrepo},
  title   = {Detecting Disturbances in Network-Coupled Dynamical Systems with Machine Learning},
  journal = {Chaos: An Interdisciplinary Journal of Nonlinear Science},
  volume  = {33},
  number  = {10},
  pages   = {103137},
  year    = {2023},
  doi     = {10.1063/5.0169237}
}

@article{michelle,
  title = {Parallel Machine Learning for Forecasting the Dynamics of Complex Networks},
  author = {Srinivasan, Keshav and Coble, Nolan and Hamlin, Joy and Antonsen, Thomas and Ott, Edward and Girvan, Michelle},
  journal = {Phys. Rev. Lett.},
  volume = {128},
  issue = {16},
  pages = {164101},
  numpages = {5},
  year = {2022},
  month = {Apr},
  publisher = {American Physical Society},
  doi = {10.1103/PhysRevLett.128.164101}
}

@article{parallelRC,
  title = {Model-Free Prediction of Large Spatiotemporally Chaotic Systems from Data: A Reservoir Computing Approach},
  author = {Pathak, Jaideep and Hunt, Brian and Girvan, Michelle and Lu, Zhixin and Ott, Edward},
  journal = {Phys. Rev. Lett.},
  volume = {120},
  issue = {2},
  pages = {024102},
  numpages = {5},
  year = {2018},
  month = {Jan},
  publisher = {American Physical Society},
  doi = {10.1103/PhysRevLett.120.024102}
}

@article{Hurley2023,
doi = {10.1088/2632-072X/ad5e55},
year = {2024},
month = {jul},
publisher = {IOP Publishing},
volume = {5},
number = {3},
pages = {035004},
author = {Hurley, L A and Restrepo, J G and Shaheen, S E},
title = {Tuning the activation function to optimize the forecast horizon of a reservoir computer},
journal = {Journal of Physics: Complexity}
}

@misc{ieee,
  author       = {{Illinois Center for a Smarter Electric Grid (ICSEG)}},
  title        = {WSCC 9-Bus System},
  howpublished = {Information Trust Institute, University of Illinois at Urbana--Champaign},
  year         = {2013},
  url = {https://icseg.iti.illinois.edu/wscc-9-bus-system/}
}

@INPROCEEDINGS{wecc,
  author={Yuan, Haoyu and Biswas, Reetam Sen and Tan, Jin and Zhang, Yingchen},
  booktitle={2020 IEEE/PES Transmission and Distribution Conference and Exposition (T\&D)}, 
  title={Developing a Reduced 240-Bus WECC Dynamic Model for Frequency Response Study of High Renewable Integration}, 
  year={2020},
  pages={1-5},
  doi={10.1109/TD39804.2020.9299666}}

@techreport{aemo2023westmurray,
  author      = {{Australian Energy Market Operator (AEMO)}},
  title       = {West Murray Zone Power System Oscillations 2020--2021},
  institution = {Australian Energy Market Operator},
  year        = {2023},
  month       = {February},
  url         = {https://aemo.com.au}
}

@techreport{NERC2025EmergingLoads,
  author      = {{North American Electric Reliability Corporation}},
  title       = {Characteristics and Risks of Emerging Large Loads},
  institution = {Large Loads Task Force (LLTF), Reliability and Security Technical Committee (RSTC)},
  year        = {2025},
  month       = {July},
  type        = {White Paper},
  address     = {Atlanta, GA},
  url         = {https://www.nerc.com/initiatives/large-loads-action-plan}
}

@article{3p8utilitydive,
  author  = {Howland, Ethan},
  title   = {PJM eyes data center, crypto reliability requirements after 3.8 GW of load trips offline},
  journal = {Utility Dive},
  year    = {2026},
  month   = {Aug},
  url     = {https://www.utilitydive.com/news/pjm-nerc-data-center-crypto-reliability-standards/827653/},
  note    = {Accessed: 2026-08-14}
}

@article{14hz,
title = {Understanding the inception of 14.7 Hz oscillations emerging from a data center},
journal = {Sustainable Energy, Grids and Networks},
volume = {43},
pages = {101735},
year = {2025},
issn = {2352-4677},
doi = {https://doi.org/10.1016/j.segan.2025.101735},
url = {https://www.sciencedirect.com/science/article/pii/S2352467725001171},
author = {Chetan Mishra and Luigi Vanfretti and Jaime Delaree and T.J. Purcell and Kevin D. Jones}
}

@inproceedings{rc2,
author = {Schrauwen, Benjamin and Verstraeten, David and Campenhout, Jan},
year = {2007},
month = {01},
pages = {471-482},
title = {An overview of reservoir computing: Theory, applications and implementations},
journal = {Proceedings of the 15th European Sympsosium on Artificial Neural Networks}
}

@article{Echo,
  author    = {Jaeger, Herbert and Haas, Harald},
  title     = {Harnessing Nonlinearity: Predicting Chaotic Systems and Saving Energy in Wireless Communication},
  journal   = {Science},
  volume    = {304},
  number    = {5667},
  pages     = {78--80},
  year      = {2004},
  publisher = {American Association for the Advancement of Science},
  doi       = {10.1126/science.1091277},
  url       = {https://www.science.org/doi/abs/10.1126/science.1091277}
}

@article{stateEst,
author = {Nazerian, Amirhossein and Nathe, Chad and Hart, Joseph and Sorrentino, Francesco},
year = {2023},
month = {10},
pages = {},
title = {Synchronizing chaos using reservoir computing},
volume = {33},
journal = {Chaos: An Interdisciplinary Journal of Nonlinear Science},
doi = {10.1063/5.0161076}
}

\end{document}